\documentclass[dvips]{article}
\usepackage{icrctc07}

\title{The VERITAS Trigger System}
\shorttitle{VERITAS Trigger}
\authors{A. Weinstein$^{1}$, for the VERITAS Collaboration$^{2}$.}
\shortauthors{A. Weinstein and et al}
\afiliations{$^1$Department of Physics and Astronomy, University of California
Los Angeles, 430 Portola Plaza, Box 951547, Los Angeles, CA 90095-1547, USA\\
$^2$For full author list see G. Maier, "Status and Performance of VERITAS",
these proceedings}
\email{amandaw@astro.ucla.edu}

\abstract{The VERITAS gamma-ray observatory, situated in southern
Arizona, is an array of four 12m diameter imaging Cherenkov
telescopes, each with a 499-pixel photomultiplier-tube camera. The
instrument is designed to detect astrophysical gamma rays at
energies above 100\/ GeV. At the low end of the VERITAS energy
range, fluctuations in the night sky background light and single
muons from cosmic-ray showers constitute significant backgrounds.
VERITAS employs a three-tier trigger system to reduce the rate of
these background events: an initial trigger which acts at the single
pixel level, a pattern trigger which acts on the relative timing and
distribution of pixel-level triggers within a single telescope
camera, and an array-level trigger which requires simultaneous
observation of an air-shower event in multiple telescopes. This
final coincidence requirement significantly reduces the rate of
background events, particularly those due to single muons. In this
paper, the implementation of all levels of the VERITAS trigger
system is discussed and their joint performance is characterized.}

\begin{document}
\maketitle


The VERITAS gamma-ray observatory, situated in southern Arizona, is
an array of four 12m diameter imaging Cherenkov telescopes, each
with a 499-pixel photomultiplier-tube camera.
The instrument is designed to detect astrophysical gamma rays with energies above 100
\/GeV.  At the low end of the VERITAS energy range, fluctuations in
the night sky background light (NSB) and single muons from
cosmic-ray showers constitute significant backgrounds, which the
three-tiered VERITAS trigger system is designed to reduce.

VERITAS has operated in a three-telescope configuration with the
full trigger system since December 2006.  A fourth telescope was
added to the array in March 2007.


\section{Level One (Pixel) Trigger}

The first tier (level one, or L1) of the VERITAS trigger system  acts
at the single pixel level.  It consists of
custom-built Constant Fraction Discriminators (CFDs), one for each
photomultplier tube (PMT) pixel in a telescope camera \cite{cfdpaper}.  The
output of the PMT is routed directly to the input of the CFD, which produces an output pulse
if the sum of the voltages from the PMT pulse and a time-delayed copy crosses a
threshold. The VERITAS CFDs are equipped with a rate feed-back (RFB) loop,
which automatically increases the effective threshold when the noise level (and
thus CFD trigger rate) rises.

Typical CFD operating parameters for the array involve a 50\/mV CFD threshold
(corresponding to approximately 4-5 photoelectrons), a 10ns output pulse, and
an RFB setting of 60mv/MHz.




\section{Level Two (Pattern) Trigger}

At each telescope, a pattern trigger system uses the relative timing and distribution
of L1 triggers within the camera to preferentially select the more
compact Cherenkov light images and reduce the rate of triggers due to random
fluctuations of night-sky background light.  The pattern trigger, hereafter
referred to as L2, is similar to that used on the Whipple 10m telescope
\cite{whipple}, but with an improved channel-to-channel timing jitter of
${<}1\mathrm{ns}$\cite{tonepaper}.  It consists of two elements, an ECL signal
splitter which copies and redirects signals from the CFDs, and 19 pattern
selection trigger (PST) modules.  The PST modules are arranged to cover
overlapping patches of the VERITAS camera and contain memory chips which can be
pre-programmed to recognize patterns of triggered pixels within the camera. The
standard pixel coincidence requirement is three adjacent pixels within a patch;
the required overlap time between adjacent CFD signals is $\sim 6$\/ns.

\section{Level Three (Array) Trigger}

\begin{figure*}
\begin{center}
\vspace{-0.4in}
\includegraphics*[width=0.6\textwidth, angle=270]{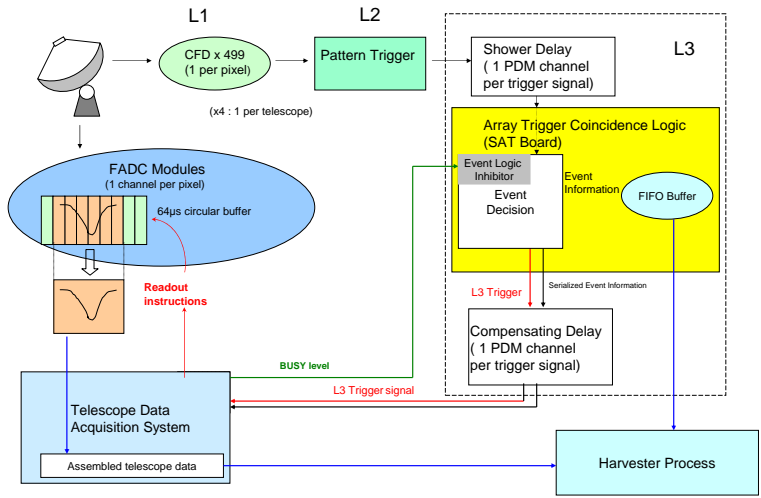}
\end{center}
\vspace{-0.5in}
\caption{Illustration of the trigger system's operation and interface with data acquisition.}\label{fig0}

\end{figure*}

As shown in Figure \ref{fig0}, the multi-telescope array
trigger (L3) receives information from the L2 trigger system at each telescope,
uses it to identify events that are consistent with simultaneous observation
of an air-shower in multiple telescopes, and provides instructions to the
telescope data acquisition systems.
The array trigger communicates with the other
systems via ECL signals; as it is centrally located, these signals are
converted and transmitted via optical fiber, using custom-built pairs of
Digital Asynchronous Transceiver modules (DATs).

A pair of custom-built VME modules, the Pulse Delay Module (PDM) and SubArray
Trigger board (SAT), together with a commercial VME GPS clock, comprise the
core of the array trigger hardware. The PDM has 32 independently programmable digital delay lines, each with a 2\/ns step
size and a 100ns-16${\mu}$s range.  The SubArray Trigger (SAT) board performs
the majority of the critical array trigger functions.  It is designed to
handle up to eight telescopes in any possible combination (subarrays).


Identification of Cherenkov shower events depends on the relative
timing of the L2 triggers.  Two main factors influence the relative
arrival times of L2 triggers at the central control building: fixed
differences in signal transmission time (due to varying optical
fiber and cable lengths) and the arrival time of the Cherenkov light
front at the different telescopes.  The first component is corrected
for exactly; the second varies as the source is tracked across the
sky, but can be approximately calculated based on the current
pointing of the telescope system.  These corrections are applied in
hardware via the delay lines of the Pulse Delay Module
(PDM) and updated on a five-second basis.

There is a residual spread in the delay-corrected L2 trigger arrival
times due to the width and curvature of the Cherenkov wavefront, variation in
the L2 trigger response with respect to image size, and timing jitter in the
various electronics components.  This spread is small (on the order of tens of
nanoseconds), allowing the SAT board to identify events by requiring L2
triggers from multiple telescopes within a fixed coincidence window (1-125ns).
The SAT converts the arrival times of the delay-corrected L2 signals into
digital time-stamps via 1.25ns resolution time-to-digital converters (TDCs) and
buffers them; the coincidence logic
algorithm, which is configured by a programmable pattern lookup table, searches
the time-stamp buffers until a programmed pattern is found
within the coincidence window.
While the lookup table can be used to suppress or privilege particular telescope combinations,
all observations to date use a simple multiplicity requirement.


Custom-built 500MS/s flash-ADC (FADC) modules (one FADC channel per
pixel) digitize the PMT signals with a memory buffer depth of 64
$\mu$s.  When the array trigger identifies an air shower event, it
sends a logical signal (the ``L3 trigger'') to each telescope,
directing the telescope data acquisition system to read out a
portion of this buffer (24 samples) for every channel.  During
readout, each data acquisition system inhibits the array trigger
coincidence logic by raising a BUSY level.  The SAT also self-vetos
for 10 \/$\mu$s after an event decision, in order to allow for L3
signal propagation to the telescopes.

Via a combination of outgoing PDM delays and internal compensation
on the SAT board, the array trigger ensures that an L3 trigger is
received at the telescope a fixed time after the corresponding L2 trigger was
produced. The data acquisition then ``looks back'' a fixed number of
FADC samples and initiates readout from that point.  This look-back
time is on the order of $3{\mu}s$ for all telescopes.  Telescopes
that do not participate in an event decision may still receive an L3
trigger (``forced readout" mode), whose timing is determined from
the timing of the participating telescope triggers.

The array trigger also tags the event with supplementary information,
including an event number, and sends it to the data acquisition system via a 48-bit
serial transmission.  This information, along with additional event
information such as a GPS timestamp, is also recorded in a FIFO. The
FIFO is polled asynchronously in software and the results are sent to
another software process, the Harvester, which binds together the
array trigger and telescope-level information into complete events.
Current polling speeds allow the array trigger to operate at rates
as high as 2kHz without data loss.

\section{Preliminary Array Trigger Characterization}

Early array trigger performance is excellent.  The array trigger
rates are extremely stable with respect to large fluctuations in the
L2 rates.  Studies of image shape parameters have already shown
\cite{tonepaper} that a multi-telescope coincidence requirement
eliminates triggers due to local muons at the 90\% level or better.
As will be shown, the array trigger is also extremely effective at
suppressing background due to NSB.

There is a large space of adjustable operating parameters for all three
levels of the trigger; full optimization studies over this entire space have not
yet been performed. However, preliminary studies were performed
\emph{in situ} to validate and characterize array performance.

\subsection{Telescope delays and coincidence window}




The time-stamps recorded by the SAT board allow us to study the pairwise L2
arrival time difference between telescopes for actual cosmic-ray showers. This
approach lets us validate the telescope delays and assess the residual spread
in L2 trigger arrival times.  We find that these distributions are centered on
zero, showing that the telescope delays have been correctly adjusted, and are
more than 99\% contained for pattern trigger separations of $\pm 25$ ns, with
negligible contributions from accidental coincidences.  Since the minimum
coincidence window width is dictated by the spread in L2 arrival times, this
behavior is consistent with the fact that the array trigger rate is
stable and independent of
coincidence window width for window sizes above 20-25ns.

\subsection{Dead-time determination and monitoring}

Accurate knowledge of the array dead-time is required in order to
determine the fluxes and spectra of astrophysical sources.  The
array trigger uses a 10MHz reference clock and a set of 32-bit scalers on the SAT
board to precisely monitor this dead-time, which is dominated by the time it takes to read out telescope information (the
average telescope readout time is $\sim 400$ $\mu$s).  As expected, the array dead-time scales linearly
with the array trigger rate, reaching $\sim$6-8\% at
150-170Hz, and 10-11\% at 225Hz.

%
As this dead-time does not scale with the L2 trigger rates, it is possible to operate the array
under conditions (such as partial moonlight) where the pattern trigger rates vary
by several orders of magnitude.

\begin{figure}[h]
\includegraphics [width=0.53\textwidth]{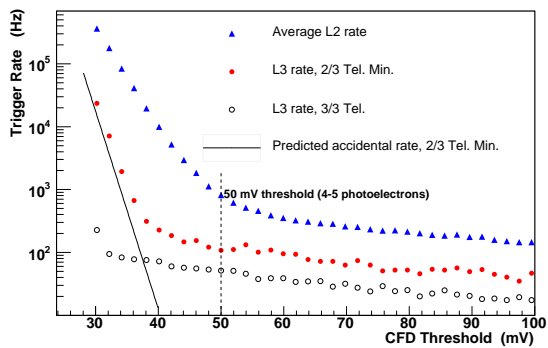}
\caption{Dependence of the L2 and L3 trigger rates on CFD threshold, for a three-telescope
array with a 50 ns coincidence window.  The L2 rate (upright
triangles) is averaged over all telescopes.  Also shown:
the L3 rates for a 2/3 (filled circles) and 3/3 (open circles)
telescope coincidence requirement, the expected accidental trigger rate for
the 2/3 requirement, as predicted from the measured L2 rates (solid line),
and the standard L1 threshold used in
array operation (dashed line).  Error bars are
commensurate with marker size.}\label{fig3}
\end{figure}


\subsection{Threshold and trigger rates}

The CFD trigger threshold, along with the other trigger operating parameters,
directly affects the energy threshold of the array.  Operating parameters must
be chosen to give the lowest possible energy threshold, while maintaining a
stable array trigger rate with an acceptable level of dead-time for a variety
of conditions.


Figure \ref{fig3} illustrates the dependence of the L2 and L3
trigger rates on CFD trigger threshold and array trigger
multiplicity requirement.
Scans of the CFD threshold were performed with
normal telescope readout disabled, so the rates shown are not
affected by the usual dead-time.  All scans were done while pointing
at a dark patch of sky near zenith, under moderate weather
conditions.

In all cases, the rates have a simple power law dependence at high
thresholds, where air-shower triggers dominate.  The L2 rates
increase rapidly in the regime dominated by accidental pixel
coincidences due to night-sky background (NSB) fluctuations.  The L3
coincidence requirement continues to suppress the NSB component of
the L3 rate, down to $\sim 40$ mV (3-4
photoelectrons) for a 2/3 multiplicity requirement and $\sim 30$ mV
(2-3 photoelectrons) for 3/3.  Below these thresholds, the array
trigger rate increases rapidly until it is saturated by accidental
coincidences.

In order to achieve stable operation with a single telescope, the
CFD threshold was set at around 70mV (6-7 photoelectrons)
\cite{tonepaper}; for array operation, a loose multiplicity
requirement of two telescopes and a CFD threshold of 50mV (4-5
photoelectrons) is used.  It is clear that a more stringent
coincidence requirement of three telescopes would allow operation at
significantly lower thresholds, but at some cost in cosmic-ray rate.




\section{Acknowledgements}

VERITAS is supported by grants from the U.S. Department of Energy, the U.S.
National Science Foundation and the Smithsonian Institution, by NSERC in
Canada, by PPARC in the U.K. and by Science Foundation Ireland.

\bibliography{icrc1144}
\bibliographystyle{plain}

\end{document}